\documentclass[12pt,preprint2]{aastex}

\usepackage{graphicx}
\usepackage{txfonts}
\usepackage{lscape}
\usepackage{aas_macros}
\usepackage{amssymb}
\usepackage{natbib}
\bibpunct{(}{)}{;}{a}{,}{,}
\usepackage{longtable}
\usepackage[hang,bf,footnotesize]{subfigure}
\usepackage[figuresright]{rotating}

\begin{document}

\title{Optical and infrared counterparts of the X-ray sources detected in the Chandra Cygnus~OB2 Legacy Survey}

\author{M. G. Guarcello\altaffilmark{1,2}, J. J. Drake\altaffilmark{2}, N. J. Wright\altaffilmark{3,2}, T. Naylor\altaffilmark{4}, E. Flaccomio\altaffilmark{1}, V. L. Kashyap\altaffilmark{2}, D. Garc\'{\i}a-Alvarez\altaffilmark{5,6,7}}

\altaffiltext{1}{INAF - Osservatorio Astronomico di Palermo, Piazza del Parlamento 1, I-90134, Palermo, Italy }
\altaffiltext{2}{Smithsonian Astrophysical Observatory, MS-67, 60 Garden Street, Cambridge, MA 02138, USA}
\altaffiltext{3}{CAR/STRI, University of Hertfordshire, College Lane, Hatfield, AL10 9AB, UK}
\altaffiltext{4}{School of Physics, University of Exeter, Stocker Road, Exeter EX4 4QL, UK}
\altaffiltext{5}{Dpto. de Astrof\'{\i}sica, Universidad de La Laguna, 38206 - La Laguna, Tenerife, Spain}
\altaffiltext{6}{Grantecan CALP, 38712 Bre{\~na} Baja, La Palma, Spain}
\altaffiltext{7}{Instituto de Astrof\'{\i}sica de Canarias, E-38205 La Laguna, Tenerife, Spain}

\begin{abstract}
The young massive OB association Cygnus~OB2, in the Cygnus~X complex, is the closest ($\sim1400\,$pc) star forming region to the Sun hosting thousands of young low mass stars and up to 1000 OB stars, among which are some of the most massive stars known in our Galaxy. This region holds great importance for several fields of modern astrophysics, such as the study of the physical properties of massive and young low-mass stars and the feedback provided by massive stars on star and planet formation process. \par
  Cygnus~OB2 has been recently observed with Chandra/ACIS-I as part of the $1.08\,$Msec Chandra Cygnus OB2 Legacy Project. This survey detected 7924 X-ray sources in a square degree area centered on Cyg~OB2. Since a proper classification and study of the observed X-ray sources also requires the analysis of their optical and infrared counterparts, we combined a large and deep set of optical and infrared catalogs available for this region with our new X-ray catalog. In this paper we describe the matching procedure and present the combined catalog containing 5703 sources. We also briefly discuss the nature of the X-ray sources with optical and infrared counterparts using their position in the color-magnitude and color-color diagrams.

\end{abstract}


\section{Introduction}
\label{intro}

The study of young stellar clusters, together with the correct classification of their stellar content, generally relies on a combination of available multi-wavelength data, from X-rays to optical and infrared. A key aspect of such studies of crowded stellar fields is the procedure adopted for merging the different datasets.  It is important to minimize the number of spurious coincidences and false negatives (i.e. sources in one waveband that fail to be matched with their real counterparts in another). A lack of accuracy and completeness in the data merging process might adversely affect source classification and the subsequent interpretation of the results. \par

  When sparse catalogs are matched, the chances of spurious coincidences are reasonably low. In these cases, simple matches based on source positions can be safely adopted.  When the source density of one or more of the catalogs is high, such that the probability of finding more than one object within the bounds of a source position uncertainty is deemed significant, the use of more complicated methods that take into account the expected multiwaveband properties of the source populations must be used. To this aim, several Maximum Likelihood methods have been proposed in the literature (e.g.: \citealp{SutherlandS1992}). \par
  
Cygnus~OB2 is the central massive OB association of the giant Cygnus X complex, with a rich population of young stars spread over an area of more than one square degree. Because of its very rich population of massive stars, Cygnus~OB2 has been described as a very young globular cluster in the Milky Way \citep{Knodlseder2000}. The census of the massive population of this association ranges from the first count by \citet{ReddishLP1967} of 300 OB members, to the estimate based on 2MASS data of more than 2600 OB stars by \citet{Knodlseder2000}. More recent studies found somewhat lower population of massive stars and identified in Cygnus~OB2 some of the most massive stars known in our Galaxy, such as O3 stars and B supergiants \citep{Walborn1973,MasseyThompson1991,ComeronPRS2002,Hanson2003,NegueruelaMHC2008}. \par 

Despite the extinction toward Cyg OB2 being high due to the intervening nebulosity associated with the Cygnus Rift \citep[roughly ranging from A$_V\sim2.5^m$ to A$_V\sim8^m$ for the optically identified members;][]{DrewGIS2008,SaleDUI2009,GuarcelloWDG2012,SaleDBF2014}, its relative proximity ($\sim1400\,$pc, \citealp{RyglBSM2012}) has made it the subject of several studies aimed at understanding its rich stellar content. Indeed, being the massive association with the largest massive star content in the proximity (i.e. within 2 kpc) of our Sun, with a massive population that has no equal in the other nearby young clusters such as the Orion Nebula Cluster, and begin also rich in pre-main sequence stars \citep{AlbaceteColomboFMS2007,WrightDrake2009,GuarcelloDWD2013}, Cyg~OB2 is also arguably the best available target to study star formation, disk evolution, and planet formation in presence of massive stars (e.g. \citealp{WrightPGD2014}). The average age of the stars in the central part of the association has been estimated to range between 3 and 5$\,$Myrs \citep{WrightDDV2010}, but several new star forming sites hosting a large fraction of very young stars still embedded in a contracting envelope or thick circumstellar disk have been discovered \citep{VinkDSW2008,WrightDDG2012,GuarcelloDWD2013}. There are also indications that some OB stars in the association are younger than $2\,$Myrs \citep{Hanson2003}, while a population of A stars found in the southern area appears to have an age between $5-7\,$Myrs \citep{DrewGIS2008}.  \par

The promise of Cyg~OB2 to be able to shed new light on the workings and products of massive star forming regions motivated a large  {\it Chandra}  X-ray Observatory $1.08\,$Msec Legacy Project \citep{Drakeinprep,WrightDGA2014}. At ages of a few million years, stars of all masses are about three and four orders of magnitude stronger X-ray emitters compared with older populations. Hard X-ray photons can penetrate many magnitudes of visual extinction and provide an effective diagnostic of youth that is free from biases resulting from accretion from a protoplanetary disk and the presence of circumstellar material. \par

The direct aim of the survey was to use the selective power of X-rays together with  the arcsecond spatial resolution of {\it Chandra} to perform a deep census of the stellar population and its X-ray properties, but with the main scientific goals of understanding the evolution of protoplanetary disks and star formation in an association approaching stellar supercluster dimensions.
The resulting {\it Chandra} catalog contains 7924 X-ray sources over an area of about one square degree centered on Cyg~OB2 \citep{WrightDGA2014}. Supplementary optical and infrared data, required  to classify the X-ray sources and follow through with the scientific objectives of our survey, have been retrieved from available public surveys (SDSS/DR8, IPHAS/DR2, UKIDSS, 2MASS) and obtained from dedicated observations with OSIRIS@GTC \citep{GuarcelloWDG2012} and Spitzer \citep{BeererKHG2010,GuarcelloDWD2013}.   \par 

A crucial step in being able to use the available multi-wavelength catalogs consists of determining which objects in one catalog correspond to sources in another. Given the large stellar density in Cyg~OB2, the depth of the OIR catalogs used, and the large foreground and background observed populations, simple nearest neighbor approaches can fail because several potential counterparts can fall within the positional uncertainty of a given source. More sophisticated approaches have employed likelihood ratio methods that seek to utilize other information than simply source position, such as comparative brightness \citep[see, e.g.,][]{Richter1975,SutherlandS1992,SmithDJB2011}.   \par

Here, we describe the matching of multi-wavelength sources  to those detected in the {\it Chandra} Cyg~OB2 survey.  
A brief overview of the X-ray, optical and infrared catalogs is presented in Sections~\ref{xraycat} and \ref{oircat}; the methods employed to cross-match objects in different catalogs are described in Section \ref{match} and the final catalog is described in Sect. \ref{catasec}. We summarize the main points of the study in Section \ref{conclusions}. \par

\section{The X-ray catalog}
\label{xraycat}

The {\it Chandra} Cyg OB2 Legacy survey design employed 36 pointings of 30$\,$ks exposure each in a 6x6 raster array heavily ($\sim 50$\%) overlapped in order to overcome the {\it Chandra} lower off-axis sensitivity and produce a relatively uniform exposure over the inner 0.5~deg$^2$ corresponding to a depth of 116 ks.  The full survey exposure was 1.08 Msec, it covered about 1 square  degree centered at 20h~33m~12s~+41~$19\arcmin$~$00\arcsec$, and was performed over a 6-weeks period from January--March 2010, employing the Advanced CCD Imaging Spectrometer (ACIS-I; \citealt{GarmireBFN2003}). 

The point source catalog was constructed using a combination of standard CIAO processing tools, source detection algorithms, and the {\it ACIS Extract} (AE; \citealt{BroosTFG2010}) software package.  In order to have an homogeneous astrometry among the various $Chandra$ pointings,{\it Chandra} astrometry was re-mapped to that of the Two Micron All Sky Survey (2MASS;  \citealt{CutriSDB2003}) using bright X-ray sources with unambiguous cross-matches to 2MASS objects.  Source detection was applied to the reduced and processed {\it Chandra} data 
in three energy bands: soft (0.5--2.0 keV), hard (2.0--7.0 keV), and broad (0.5--7.0 keV) using different algorithms: An enhanced version of the CIAO tool {\sc wavdetect} that performs source detection on multiple non-aligned X-ray observations, detecting sources that may not be detected in the individual observations, and {\sc pwdetect} \citep{DamianiMMS1997}.
This process was augmented by several hundred sources from lists of known Cyg OB2 members, including O and B-type stars \citep{Wrightinprep_obpopulation} and young A-type stars  \citep{DrewGIS2008}, creating a total of 13,041 source candidates.  

Candidate source photometric extraction and validation was performed using AE in an iterative fashion, whereby validated sources were excluded from regions used for background estimation, followed by a repeat of the AE extraction and validation.  Owing to the overlapping source and background regions in the most crowded areas of the survey, several iterations of this process were required.  The resulting X-ray catalog contains 7924 verified sources, 
the vast majority of which were observed at least 4 times in overlapping tiles, and detected within 4 arcmin of the telescope optical axis at least once.  The source positional uncertainty is typically $<0.5\arcsec$ and we estimate a 90\% completeness for stellar X-ray sources down to X-ray luminosity of $7\times10^{29}$~erg~cm$^2$~s$^{-1}$ in the central 0.5 square degrees.  A full description of the catalog construction is presented by \citet{WrightDGA2014}, while an assessment of the catalog contents and sensitivity are discussed by \citet{Wrightinprep_sensitivity}.

\section{The optical-infrared catalog}
\label{oircat}

The optical-infrared (OIR) catalog used in this work contains photometric data retrieved from several publicly available catalogs:

\begin{itemize} 
\item the optical catalog in $r,\,i,\,z$ bands (65349 sources) obtained from observations with the Optical System for Imaging and low Resolution Integrated Spectroscopy (OSIRIS), mounted on the $10.4\,$m Gran Telescopio CANARIAS (GTC) of the Spanish Observatorio del Roque de los Muchachos in La Palma \citep{CepaAEG2000} compiled by \citet{GuarcelloWDG2012}; 

\item the second release of the optical catalog in $r^{\prime},\,i^{\prime},\,H\alpha$ bands (24072 sources) obtained from observations with the Wide Field Camera (WFC) on the $2.5\,$m Isaac Newton Telescope (INT) for the INT Photometric H$\alpha$ Survey (IPHAS, \citealt{DrewGIA2005,BarentsenFDG2014}); 

\item the SDSS catalog (eighth data release, DR8, 27531 sources, \citealp{AiharaAAA2011}) in $u,\,g,\,r,\,i,\,z$ bands;

\item the UKIDSS/GPS catalog in the $JHK$ bands \citep{HewettWLH2006,LucasHLS2008}, containing 273473 sources, from observations taken with the Wide Field Camera (WFCAM, \citealp{Casali2007}) on the United Kingdom InfraRed Telescope (UKIRT), compiled adopting a new photometric procedure \citep{KingNBG2013} based on the UKIDSS images \citep{DyeWHC2012}; 

\item the 2MASS/PSC catalog  in $JHK$ (\citealp{CutriSDB2003}, 43485 sources);

\item the catalog in the Spitzer/IRAC $3.6,\,4.5,\,$ $5.8,\,8.0\,\mu$m and MIPS $24\mu$m bands (149381 sources) from the Spitzer Legacy Survey of the Cygnus~X region Spitzer \citep{BeererKHG2010}. 

\end{itemize}

As described in \citet{GuarcelloDWD2013}, these catalogs have been combined into a large OIR catalog containing 329514 sources. The matching procedure was divided into three steps. First, a combined optical catalog was produced by matching the OSIRIS, IPHAS, and SDSS catalogs. Second, an infrared catalog was created by matching UKIDSS, 2MASS and Spitzer data. In the last step, the two catalogs were merged into an unique OIR catalog. All the data used here, except those from OSIRIS, are available over the entire area surveyed with Chandra/ACIS-I. The OSIRIS data are only available in a central $40^{\prime}\times 40^{\prime}$ field. \par
  The OIR catalog includes stars associated with Cygnus~OB2 down to very low masses.  Assuming a distance of $1.4\pm0.08\,$kpc \citep{RyglBSM2012}, an average extinction $A_V=4.3^m$ \citep{GuarcelloWDG2012}, and adopting the isochrones of \citet{SiessDF2000}, we can estimate that we have good quality optical and infrared data for members down to $0.2\,$M$_{\odot}$, allowing us an unprecedentedly deep and complete study of the population of Cygnus~OB2.

\section{The adopted matching procedures}
\label{match}

The X-ray sources in our survey need to be classified also according to their OIR properties \citep{Kashyapinprep}.  Erroneous matches between the OIR and X-ray catalogs will result in wrong classifications, affecting the scientific outcome of our survey. For this reason, particular attention must be given to how the OIR and X-ray catalogs are merged. \par
  A simple matching procedure based on the positions of the sources and using a fixed matching radius (i.e. considering as real counterparts the OIR and X-ray pairs with a separation smaller than a given threshold) is unsuitable to our case for two reasons. First, the Point Spread Function (PSF) of the {\it Chandra} mirrors increases in size with increasing off-axis angle. For this reason, the positional accuracy of the X-ray sources is not constant across the field. Second, while the optical data are dominated by the foreground stellar population, and the infrared data by the background sources, in both cases with an approximately uniform spatial distribution, most of the X-ray sources with OIR counterparts are expected to be associated with Cygnus~OB2 and clustered at the locations of the various subclusters of the association. The density of the sources not associated with the X-ray population (the {\it uncorrelated} population) is high, and any attempt at matching the OIR and X-ray catalogs using only positional information will inevitably result in large numbers of spurious matches.  It is necessary, then, to use a more sophisticated approach. \par

One method used successfully in similarly challenging matching procedures is based on Maximum Likelihood (ML, \citealp{SutherlandS1992}) approaches that takes into account both the spatial separation between the different catalog sources (OIR and X-ray in our case) and how the magnitude of the OIR sources compare with those expected for the real OIR counterparts of the X-ray sources (the {\it correlated} population). Several ML methods have been used in the literature (i.e. \citealp{TaylorMEB2005,GilmourGAB2007,RumbaughKGL2012}). In this work, rather than rely on a single matching procedure, we adopt three different methods. The final OIR-X-ray catalog will contain all the pairs matched with each of the three methods, with the subsample of the most reliable matches properly tagged.

\subsection{Modified \citet{SmithDJB2011} procedure}
\label{smithmatch}

One of the methods that we adopted is defined in \citet{SmithDJB2011}, slightly modified in order to optimize it for our specific multi-wavelength case. In this approach, the probability that a given OIR source is the correct counterpart of a nearby X-ray source is calculated starting from the following likelihood ratio:
\begin{equation}
LR=\frac{q \left( m \right) f \left( r \right)}{n \left( m \right)}
\label{like_eq}
\end{equation}
In this definition, $f\left( r \right)$ is the radial distribution function of the separations between OIR and X-ray pairs as a function of the positional error:
\begin{equation}
f \left( r \right) = \frac{1}{2 \pi \sigma_{pos}^2} exp\left( \frac{-r^2}{2\sigma_{pos}^2} \right)
\label{fr_eq}
\end{equation}
Here, $r$ is the positional offset between OIR and X-ray sources, and $\sigma_{pos}$ the positional uncertainties, calculated adding in quadrature the OIR and X-ray positional uncertainty. The quantities $q(m)$ and $n(m)$, i.e. the magnitude probability distributions of the correlated OIR sources and the observed magnitude probability distribution of all the OIR sources in the $m$ band, respectively, are described in the next sections.

\subsubsection{The observed magnitude distributions}
\label{qm}

In Eq. \ref{like_eq} $q\left( m \right)$ and $n\left( m \right)$ are the probabilities to observe, respectively, a correlated and a generic OIR source with magnitude $m$. The main difference between the method we used and that defined in \citet{SmithDJB2011} is that the latter method is applied using one optical catalog. Our multi-wavelength catalog contains data from various optical and infrared catalogs, and most of the sources lack a detection in one or more of them. For instance, highly embedded or extinguished objects in the background, or even associated with the most obscured regions of Cygnus~OB2, often lack optical counterparts. 

For this reason, we seek to use all the OIR information available in order to improve the completeness of the final OIR+X-ray catalog. We calculated, then, $q\left( m \right)$ and $n\left( m \right)$ for each band available in our OIR catalog. We also defined for each OIR source a representative band, which is the first one available and with an error smaller than $0.1^m$ proceeding from shorter to longer wavelengths, starting from the $r$ band. In the optical bands we used preferentially the OSIRIS photometry. The IPHAS photometry has been used when OSIRIS data are not available, and SDSS photometry when there are no other optical data. Our main catalog in the near infrared is UKIDSS, while 2MASS data are used when UKIDSS data are not available or of bad quality. This priority among the available bands has been arbitrary chosen, after having verified that the chosen order was not affecting our results, since it was possible to define a representative band for almost all the bands. For the vast majority of sources in our OIR catalog, the representative bands are the OSIRIS $r$ or the UKIDSS $J$ bands.  
  
To obtain  $n\left( m \right)$, we first calculated the observed magnitude distributions in each band of our catalog, sampled in bins of $0.25^m$ of width. The probability $n\left( m \right)$ for a given OIR source is then given by the fraction of sources observed in our OIR catalog in the representative band in the same magnitude bin, normalized by the total area of the survey. 

\subsubsection{The correlated magnitude distributions}
\label{nm}

The calculation of $q\left( m \right)$ for each OIR source is more complicated, since it requires the computation of the magnitude distribution of the expected correlated population, after considering and removing the contribution from the uncorrelated population. Following \citet{SmithDJB2011}, the initial approximation of the expected magnitude distribution of the correlated population is obtained from all the OIR sources closer than $10^{\prime \prime}$ to the X-ray sources (hereafter the {\it nearby} population). Even at large off-axis angles, this matching radius is significantly larger than the propagated positional uncertainty, resulting in a selection of 78182 {\it nearby} OIR sources for the 7924 X-ray sources. The expected magnitude distribution of the correlated population is obtained from that of the {\it nearby} population, by subtracting in each magnitude bin the number of uncorrelated sources expected to match the positions the X-ray sources and falling in the given bin of magnitude:

\begin{equation}
q \left( m \right) = nearby\left( m \right) - N\left( m \right) \times N_{x} \times \frac{\Delta_{match}}{\Delta_{tot}},
\label{nm_eq}
\end{equation}

where $nearby\left( m \right)$ is the magnitude distribution of the {\it nearby} sources, $N_{x}$ is the total number of X-ray sources (7924), $\Delta_{match}$ is the matching area with a radius of 10$^{\prime\prime}$, $\Delta_{tot}$ is the total area of our survey (1 square degree), and $N\left(m \right)$ is the observed magnitude distribution $n\left( m \right)\times \Delta_{tot}$. By using this formula, we are also assuming that the uncorrelated sources are uniformly distributed in the survey area, which may be incorrect in case of not uniform extinction such as in Cyg~OB2.   \par
\begin{figure*}[!t]
        \centering
        \includegraphics[width=15cm]{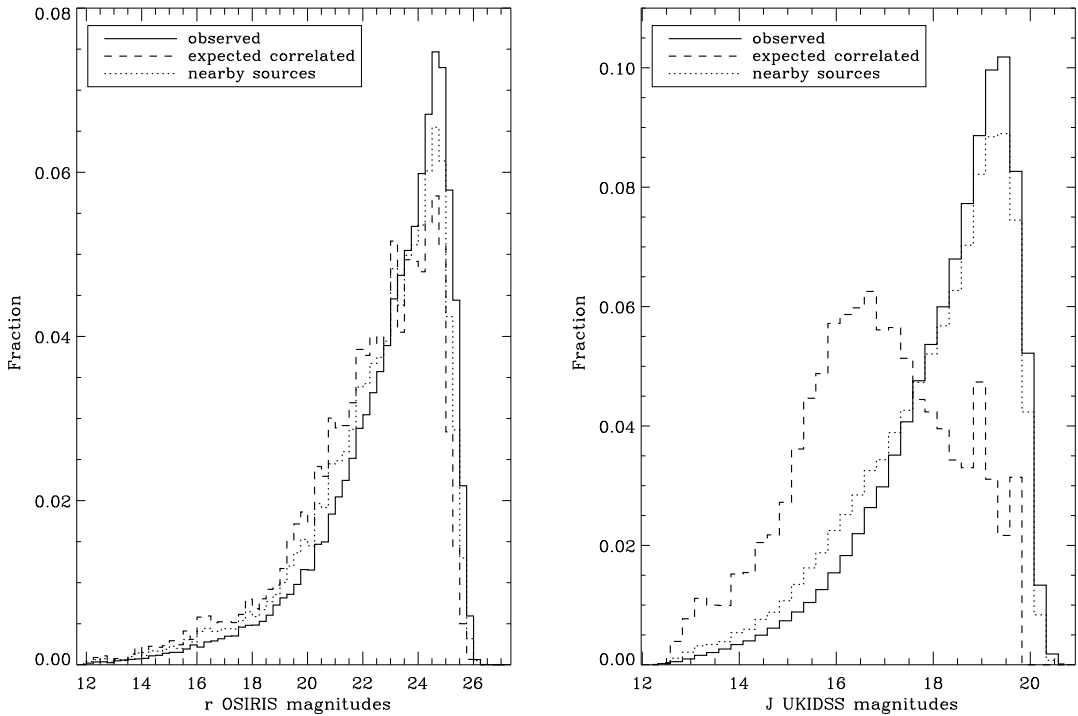}
        \caption{Magnitude distributions in the OSIRIS $r$ (left panel) and UKIDSS $J$ (right panel) bands for the entire OIR catalog (solid histogram), the {\it nearby sources} (dotted histogram), and the expected correlated distributions (dashed histograms) using Eq. \ref{nm_eq} (SM method)}
        \label{magdis_smi}
        \end{figure*}

  Fig. \ref{magdis_smi} shows the magnitude distributions in the OSIRIS $r$ and UKIDSS $J$ bands for the entire OIR catalog, the {\it nearby} sources, and the expected correlated population. In the optical band there is not much difference between these distributions, not even in the faint part. This may indicate that this method is not very effective in removing fortuitous coincidences between X-ray sources and faint background optical sources. The correction used to remove the uncorrelated population has been more effective in the $J$ band, as demonstrated by the difference between the total distribution, centered at $J=19.5^m$, and the expected correlated distribution, centered in the range $16^m<J<17^m$.

\subsubsection{Reliability associated with an OIR+X-ray pair}
\label{rij}

Once we have calculated $f\left( r \right)$, $q\left( m \right)$, and $n\left( m \right)$, we can obtain $LR$ for each pair of X-ray and OIR sources from Eq~\ref{like_eq}. Using the value of $LR$, we can assign a probability that a given OIR source is the real counterpart of the nearby X-ray source, and compare it with a chosen threshold (see Sect. \ref{results}). This probability is calculated comparing the observed $LR$ value with a distribution of $LR$ values obtained from 200000 test X-ray sources uniformly distributed across the field. The use of uniform spatial distribution of the X-ray sources is an acceptable approximation since the OIR catalog, dominated by background NIR sources, has a nearly uniform spatial density. These test sources were matched with the OIR catalog, obtaining a distribution of simulated $LR$ values from more than 70000 pairs (the exponential form in $f\left( r \right)$ cuts any match between sources more distant than few arcseconds). The {\it reliability} associated with each match between the X-ray and OIR sources in our catalog, which is, by definition, the probability that the given OIR source is the real counterpart of the nearby X-ray source, is then calculated as: 
\begin{equation}
R_{ij}=1-\frac{N_{gt}}{N_{sim}}
\label{rij_eq}
\end{equation}
where $R_{ij}$ is the reliability that the OIR source $i$ is the real counterpart of the X-ray source $j$; $N_{sim}$ is the number of simulated $LR$ values; $N_{gt}$ is the number of simulated $LR$ values larger than the one observed between the $ij$ pair:
\begin{equation}
N_{gt}=N \left( LR_{simul} > LR_{ij} \right)
\label{Ngt_eq}
\end{equation}
In this way each $ij$ pair (i.e. each pair of X-ray and OIR sources) has an associated probability that the OIR source is the real counterpart of the X-ray source. 

\subsubsection{Match results}
\label{results}

The last step consists in assigning a probability cut-off, i.e. to decide what is the minimum reliability that identifies real matches. This was performed by studying how the number of spurious matches out of the total number of matches increases with decreasing the cut-off. To obtain the number of spurious matches, we repeated the matching procedure after ``randomizing'' our X-ray catalog, i.e. applying rigid translations of $1^{\prime}$ to the X-ray sources four times, each time with a different combination of positive and negative rigid translations in RA and DEC. The number of expected spurious matches corresponding to given test values of the cut-off is the mean of the number of matches obtained with these four ``randomized'' X-ray catalogs. We then fixed our cut-off value as the one for which the ratio of spurious to total matches is $\sim 10\%$ (corresponding to $R_{cut-off}=0.95$). 

\begin{figure}[!t]
        \centering
        \includegraphics[width=8cm]{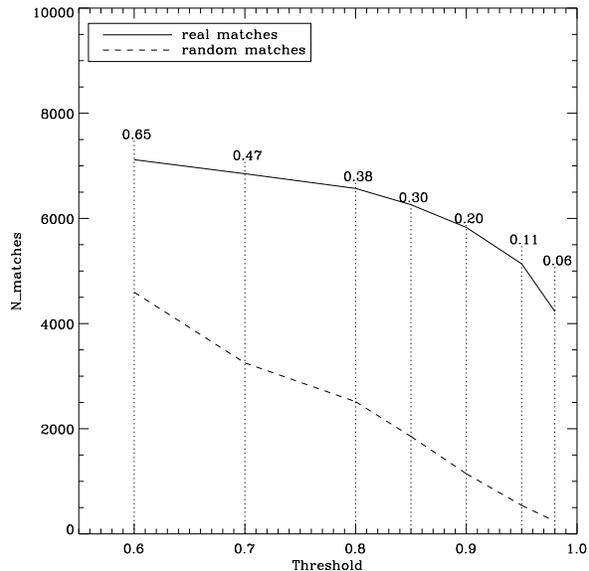}
        \caption{Number of real (solid line) and spurious (dashed line) matches obtained with increasing the reliability threshold in the SM method. The numbers over the solid line show how the fraction of spurious matches decreases with increasing the threshold}
        \label{thresh_diag}
        \end{figure}

Fig. \ref{thresh_diag} shows how the number of total and spurious matches, together with their ratio, vary with the test thresholds. With the chosen cut-off of 0.95, we matched 5180 pairs, with 4946 single matches. Hereafter, this method is called the {\it SM method}.  
 
\subsection{Matching procedure with the {\it correlated} population from an accurate position match.}
\label{pmmatch}

The second matching procedure is based on a different definition of the {\it correlated} population. The definitions of $LR$ (Eq. \ref{like_eq}), of the observed magnitude distributions $n\left( m \right)$, as well as that of the reliability (Eqs. \ref{rij_eq} and \ref{Ngt_eq}) and the procedure to define the threshold (Sect. \ref{results}) are the same as those adopted in the previous sections. 

In Sect. \ref{nm} we calculated the magnitude distribution of the expected {\it correlated} population starting from a position match between the X-ray and OIR catalog with a large matching radius ($10^{\prime \prime}$), and then we used a statistical approach to remove from the {\it nearby} population the expected contribution of the {\it uncorrelated} OIR sources. This correction was necessary, as proved by the very large number of $nearby$ sources found (78182) and the risk that the SM method may not be very effective in removing spurious coincidences between X-ray and faint optical sources in the background. However, since the chances that OIR sources nearby X-ray positions are real counterparts decrease with increasing separation, a different estimate of the {\it correlated} population can be found with a nearest-neighbor match using suitable small matching radii estimated with a detailed statistical analysis of real and spurious matches. 
  
\subsubsection{The {\it correlated} population from accurate nearest-neighbor match}
\label{pmradius}

In this procedure, we obtained a {\it correlated} population from which we derived $q \left( m \right)$ from an accurate nearest-neighbor match. The matching radius used in this procedure must take into account the degradation of the PSF in the X-ray images at increasing off-axis angles and the photon statistics of the X-ray source. In order to do that, we used an individual matching radius for each X-ray source which is proportional to the X-ray positional uncertainty, the latter calculated as in \citet{KimKWG2007}:
\begin{equation}
r_{match}=A \times \sigma_{pos}\\
\label{matchrad}
\end{equation}
\begin{equation}
log \left(\sigma_{pos} \right) = 0.1137 \Theta -0.46 log \left( C \right) -0.2398\\
\label{err1}
\end{equation}
\begin{equation}
log \left(\sigma_{pos} \right) = 0.1031 \Theta -0.1945 log \left( C \right) -0.8034
\label{err2}
\end{equation}
where $r_{match}$ is the individual matching radius, $A$ is a coefficient to be evaluated, $\sigma_{pos}$ is the positional uncertainty, $\Theta$ is the off-axis angle of the X-ray sources and $C$ is the net number of counts. Following \citet{KimKWG2007}, Eq.~\ref{err1} is applied to sources with less than 133 counts; Eq. \ref{err2} to brighter sources. 

  Given the different depth and spatial distribution of the optical, $JHK$ (UKIDSS and 2MASS), and Spitzer catalogs, we decided to perform the nearest-neighbor match for each of these three catalogs separately and then to merge the results. The crucial step here is to estimate the coefficients $A$ for the three catalogs, together with a minimum allowed matching radius ($r_{min}$). The use of $r_{min}$ is necessary since the positional errors in the center of the ACIS field are very small, resulting in unacceptably low matching radii. 
  
The procedure adopted to calculate these parameters is similar to that defined in Sect.~\ref{results}, i.e. by comparing the ratio of the numbers of spurious coincidences to that of the total matches obtained with increasing test values of $A$ and $r_{min}$. The spurious coincidences are calculated by matching the OIR catalogs with ``randomized'' X-ray catalogs (as in Sect. \ref{results}); while the total number of matches by combining the OIR catalogs with the ``real'' X-ray catalog (i.e. with no positional offset added). We first evaluated $A$ and then $r_{min}$, in both cases as the largest test values at which the spurious matches reached $\sim 10\%$ of the real matches. Table \ref{match_tb} lists the values of $A$ and $r_{min}$ found for the optical, $JHK$ and {\it Spitzer} catalogs, together with the total number of matches. The catalog of the expected {\it correlated} sources obtained by merging the results of these three nearest-neighbor matches numbers 5820 sources, many less than the {\it nearby sources} defined in Sect. \ref{nm} and comparable to the final number of the OIR+X-ray pairs matched in the merged catalog (Sect. \ref{catasec}). This catalog has been used to define the magnitude distribution $q \left( m \right)$ used in Eq. \ref{like_eq}. 

\begin{figure*}[!t]
        \centering
        \includegraphics[width=15cm]{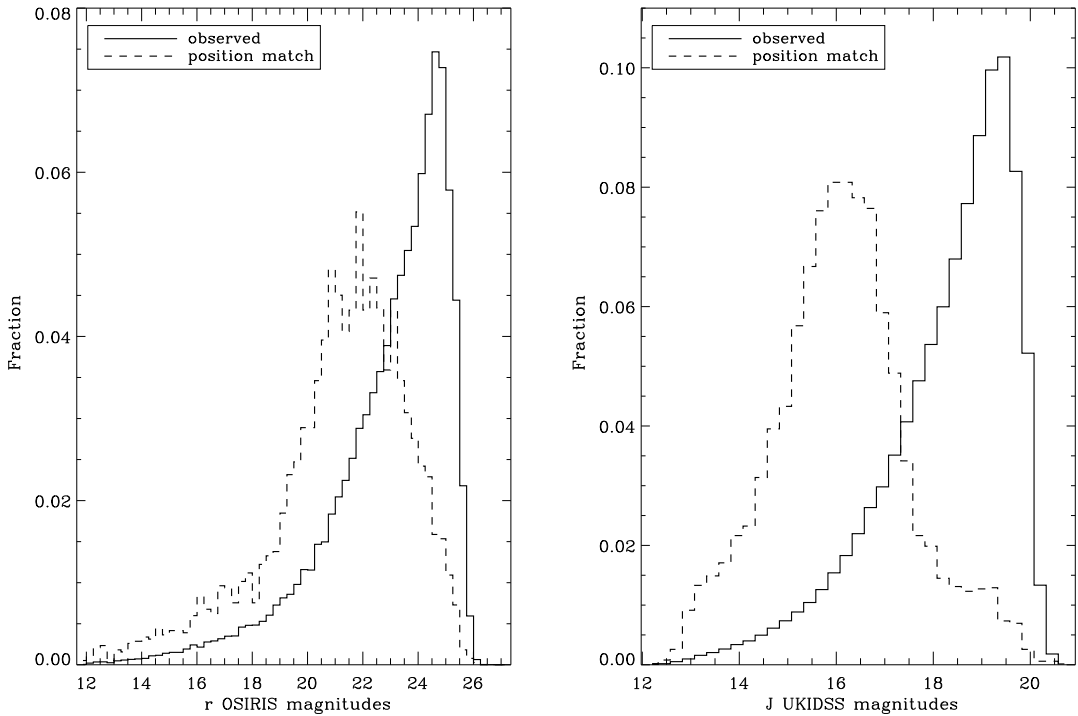}
        \caption{Magnitude distributions in the OSIRIS $r$ (left panel) and UKIDSS $J$ (right panel) bands for the entire OIR catalog (solid histogram), and for the OIR sources matched with the X-ray sources using an individual matching radius as described in Sect. \ref{pmradius}, i.e. in the PM method (dashed histogram)}
        \label{magdis_pm}
        \end{figure*}

	\begin{table}[]
        \centering
        \caption {Results of the close-neighbor matches}
        \vspace{0.5cm}
        \begin{tabular}{cccc}
        \hline
        \hline
        OIR catalog & $r_{min}$ & $A $ & N. matches \\
       \hline
	Optical		& $0.6^{\prime \prime}$ & 1.4 & 4917 \\
	$JHK$		& $0.5^{\prime \prime}$ & 0.9 & 5025 \\
	Spitzer		& $0.5^{\prime \prime}$ & 1.3 & 5278 \\
        \hline
        \hline
        \multicolumn{4}{l}{} 
        \end{tabular}
        \label{match_tb}
        \end{table}
Fig. \ref{magdis_pm} shows the distributions of the OSIRIS $r$ and UKIDSS $J$ magnitudes for all the OIR sources and for the expected {\it correlated} population found with the accurate nearest-neighbor match described here.  Comparison with Fig.~ \ref{magdis_smi} reveals that the magnitude distributions obtained with this method are shifted for brighter magnitudes: The $r$ distribution of the {\it correlated} population in Fig.~\ref{magdis_pm} is centered in the range $20^m<r<22^m$, while in Fig.~\ref{magdis_smi} it is peaked at about $r\sim24^m$ similar to the distribution of the total OIR catalog. The effect in the $J$ distribution is smaller. This indicates that this method is more effective than the SM method in removing candidate spurious matches with faint sources. 

The rest of the procedure (i.e.\ the calculation of the reliability and the threshold for reliable matches, equal to 0.96) was performed as described in Sects.~\ref{rij} and \ref{results}, and resulted in 5210 matches (4933 single matches). Hereafter, this method is called the {\it PM method}.

\subsection{Naylor et al.~2013 Bayesian method}
\label{bayesmeth}

 A Bayesian technique to cross correlate X-ray catalogs with deep infrared data has been developed by \cite{NaylorBF2013}, as an extension of the techniques defined in \citet{SutherlandS1992}.  This method has several similarities with the maximum likelihood methods described above.  As in those methods, the magnitude of the candidate OIR counterpart is compared with an expected magnitude distribution for the correlated population to reduce the chances of matching the X-ray sources with an uncorrelated background or foreground source. \par
 
  We defer a description of the method to \cite{NaylorBF2013}, though note that it brings two significant improvements to existing match criteria such as those we used to define our ML methods. First it has a more sophisticated method of estimating the magnitude distribution of the correlated population, which allows for the fact that the presence of counterparts in the error circles means they are more crowded than the field. The second improvement lies in accounting simultaneously for all the OIR sources close to an X-ray position to calculate the probability that each of them is the real counterpart of an X-ray source. This is not the case with the methods described above, where the reliability is calculated independently from the values obtained for the other nearby OIR sources. The main gain of the more sophisticated approach is in reducing chances of multiple source matches. \par
  
We applied the method using the $i$-band, $K$-band, and [3.6] photometry, where the number of available sources with good photometry is larger, merging the various catalogs available for each band using the hierarchy described in Section \ref{qm}. Since the matching method requires a magnitude for each star, we removed from the catalog all objects which did not have magnitudes in the band in question, or a magnitude whose uncertainty was greater than 0.3$^m$.\par

  The first stage of the matching process is to check the model of the X-ray error circles by comparing the distribution of stars around all the X-ray positions with that predicted by a model consisting of a uniform background and a set of counterparts. The latter were initially assumed to have a Gaussian distribution about the X-ray positions with the radius given by the \citet{KimKWG2007} model. We found the data were best fitted by multiplying the radius of the error circles by 0.6 and adding in quadrature a position and X-ray flux independent uncertainty of 0.2$\arcsec$. The combination suggests the \citet{KimKWG2007} radii over-estimate the positional uncertainty in our data by perhaps 20 percent. We also explored systematic offsets in the data, and found then best fit corresponded to a shift in RA of 0.02$\arcsec$ with no shift in declination. \par
  
The remainder of the matching process proceeded as described in \cite{NaylorBF2013}, and resulted in a list of all stars which had a likelihood of being a counterpart greater than 0.05. The counterpart probabilities are presented along with those from the other methods in Table \ref{catalog_tb}. We {included in the list of counterparts} only those stars whose likelihoods exceeded 0.8. We can estimate the contamination in this sample by summing all the likelihoods that a given star is not the counterpart, and then dividing this by the number of counterparts. For both the $K$ and $i$ band samples with likelihoods greater than 0.8 this gives a contamination rate of about 2\%. \par

  The total number of sources matched with this method is 5157. Considering only the single matches, the number of sources matched by this method is similar to those matched by the two ML methods: 4933 sources with the PM method, 4946 with SM, and finally 4958 with this method, called hereafter the {\it NBF method}.

\section{The final OIR+X-ray catalog}
\label{catasec}

\subsection{Reliable matches}
\label{reliablematches}

\begin{figure}[!t]
        \begin{center}
        \includegraphics[width=8.5cm]{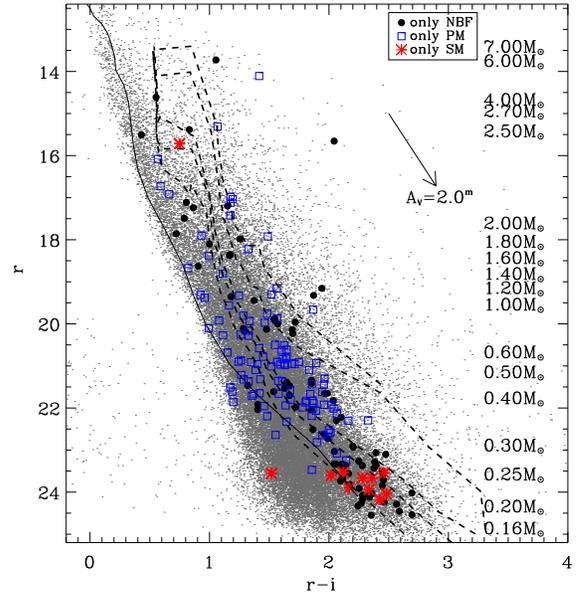}
        \caption{$r$ vs. $r-i$ diagrams with all the sources with good optical photometry marked with gray dots, the isochrones with age $<10\,$Myrs and the ZAMS plotted using the distance and extinction as in \citet{GuarcelloWDG2012}. The reddening vector and the masses corresponding to the $3.5\, Myrs$ isochrone are also shown. Different symbols and colors mark the X-ray sources matched with just one method. }
        \label{colmag}
        \end{center}
        \end{figure}

The catalog merged from all the above methods numbers 5703 OIR+X-ray pairs matched with at least one method: 4643 are matched with all three methods, 558 with two, and 502 with just one. Among the matched sources, there are 5398 single matches and 305 multiple matches. Fig. \ref{colmag} shows the $r$~vs.~$r-i$ diagram of all the optical sources with good photometry within the Chandra Cygnus~OB2 Legacy field. The conditions for good photometry are defined in \citet{GuarcelloWDG2012}: In short, they require small errors ($\sigma_{r}<0.1^m$ and $\sigma_{r-i}<0.15^m$) in at least one of the three optical catalogs we have used. Also shown in Fig.~\ref{colmag} are the isochrones and ZAMS from \citet{SiessDF2000},  plotted using a distance of $1.4\pm0.08\,$kpc \citep{RyglBSM2012}, with the extinction $A_V=4.3^m$ for the isochrones and $A_V=1^m$ for the ZAMS \citep{GuarcelloWDG2012}. In the right side of the diagram are marked the $r$ magnitudes of $3.5\,$Myrs old stars with $A_V=4.3^m$ and $d=1.4\pm0.08\,$kpc with different masses. The extinction vector is obtained from the extinction law of \citet{Donnell1994}. The sources matched with only one method are overplotted with different symbols and colors, as explained in the legend. The sources matched only with the NBF method lie mainly in the cluster locus (i.e.\  in this diagram the area between the isochrones), spanning all the magnitude range but with a larger fraction of faint sources. Following the reddening vector, these sources can be stars associated with Cyg~OB2 suffering large extinction, or low mass members observed during a period of high X-ray activity. The analysis of their X-ray spectra and light curves will help us to better classify them \citep{Flaccomioinprep1}. The sources matched only with the PM method populate mainly the region of the diagram with intermediate magnitudes. Most of them are compatible with the cluster locus. The sources matched only with the SM method lie only in the faint end of the cluster locus. The fact that all the sources matched only by the SM method are faint is likely a consequence of the lower effectiveness of the SM method in removing spurious matches with faint optical uncorrelated sources (Fig. \ref{magdis_smi}). These samples show the same properties in all the optical and infrared color-magnitude diagrams. \par

\begin{figure}[!t]
        \centering
        \includegraphics[width=8.5cm]{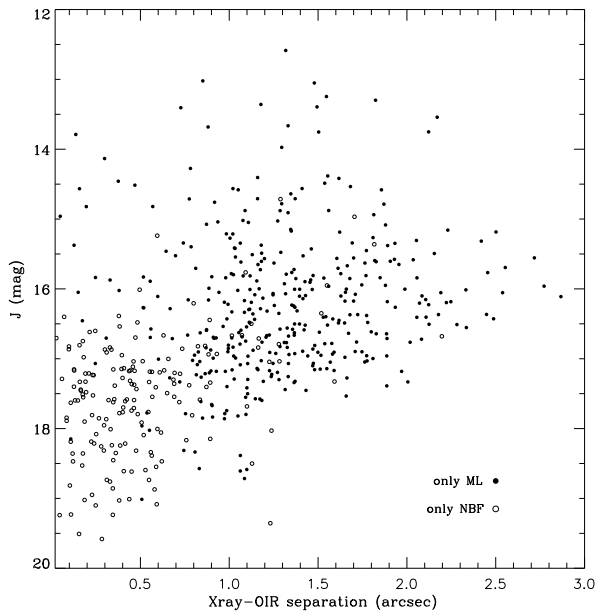}
        \caption{$J$ magnitude of the OIR counterparts vs. the separation between the OIR and the X-ray counterparts for the sources matched only with the NBF or the ML methods.}
        \label{checktay}
        \end{figure}

  Fig.~\ref{checktay} shows the $J$ magnitudes of the OIR counterparts of the matched sources vs. their separation from the X-ray sources, for the pairs matched only with the NBF method and those matched only with the ML methods. In this diagram the differences between these two samples are evident: the stars matched only with the NBF method are systematically fainter and closer, indicating that the ML methods (mainly PM) has been very conservative in removing candidate spurious coincidences with close faint counterparts. On the other hand, the ML methods are more effective in matching OIR counterparts with intermediate and bright magnitudes and at large separations (i.e. larger than 1$^{\prime \prime}$), which are plausible given the size of the ACIS PSF.  \par

Table \ref{catalog_tb} shows a description of the merged OIR and X-ray catalog. The catalog contains 79 columns: The first 55 show the optical and infrared photometry of the OIR counterpart, followed by seven columns with some of the X-ray properties of the X-ray counterpart from \citet{WrightDGA2014}. The columns from the 63th to the 72th contain tags indicating the matching procedure that matched the two counterparts, with the related probabilities, while the last columns information on the multiple matches. As explained earlier, the 305 multiple matches are kept in the merged catalog. Different scientific uses of the catalog require different treatment of the multiple matches. For instance, studies based on stellar positions may simply consider the multiple matches as a single entry at a given position; or studies based on the photometric properties would require to discard the multiple matches. For those cases where one of the matching counterparts in the multiple matches must be chosen, we provide in the catalog a column indicating the ``best counterpart''. Users must use this column and deal with multiple matches with caution.  Given the evidence that the SM method is not efficient in removing spurious coincidences between the X-ray and faint optical sources, and the fact that the expected magnitude distribution of the correlated stellar population is calculated with a statistical approach which is not fully suitable for our survey, we consider as ``reliable matches'' the 5619 sources matched by either the NBF (the NBF sample) or PM (the ML sample, i.e. removing the sources matched only by the SM method) method. \par

        \begin{table*}[]
        \centering
        \caption {List and description of the columns of the optical, infrared and X-ray catalog of the Chandra Cygnus~OB2 Legacy Survey}
        \vspace{0.5cm}
        \begin{tabular}{ll}
        \hline
        \hline
        Columns & Description \\
       \hline
  1 & Sequential ID \\
  2-3& Right Ascension and Declination in J2000 (projected on 2MASS) \\
  4-9& OSIRIS photometry in $riz$ OSIRIS bands \\
  10-15& IPHAS photometry in $riH_{\alpha}$ IPHAS bands \\
  16-25& SDSS photometry in $ugriz$ SDSS bands \\
  26-31& 2MASS photometry in $JHK$ 2MASS bands \\
  32-35& 2MASS quality flags {\it ph\_qual, rd\_flg, bl\_flg, cc\_flg} \\
  36-45& UKIDSS photometry in $JHK$ plus the $J-K$ and $H-K$ colors \\
  46-53& IRAC photometry at $3.6,\, 4.5,\, 5.8,\, 8.0\, \mu m$ \\
  54-55& MIPS photometry in $24\, \mu m$ \\
  56& X-ray ID from \citet{WrightDGA2014}  \\
  57& exposure time of the X-ray source in sec \\
  58-60& net, total, and background counts of the X-ray source \\
  61& median photon energy observed \\
  62& separation between the X-ray and the OIR counterparts in arcsec \\
  63& tag equal to 1 if the pair is merged by the PM method \\
  64& tag equal to 1 if the pair is merged by the SM method \\
  65& tag equal to 1 if the pair is merged by the NBF method using $i$ \\
  66& tag equal to 1 if the pair is merged by the NBF method using $K$ \\
  67& tag equal to 1 if the pair is merged by the NBF method using [3.6] \\
  68& probability that OIR and X-ray source are real counterparts from the PM method \\
  69& probability that OIR and X-ray source are real counterparts from the SM method \\
  70& probability that OIR and X-ray source are real counterparts from the NBF method using $i$ \\
  71& probability that OIR and X-ray source are real counterparts from the NBF method using $K$ \\
  72& probability that OIR and X-ray source are real counterparts from the NBF method using [3.6] \\
  73& multiple OSIRIS-IPHAS matches (0 if a single match) \\
  74& multiple UKIDSS-2MASS matches (0 if a single match) \\
  75& multiple NIR-Optical matches (0 if a single match) \\
  76& multiple NIR+OPT-Spitzer matches (0 if a single match) \\
  77& multiple OIR-X-ray matches (0 if a single match) \\
  78& equal to 1 if it is a preferred counterpart (always 1 in single matches) \\
  79& equal to 1 if it is a reliable match \\

	\hline
        \hline
        \multicolumn{2}{l}{} 
        \end{tabular}
        \label{catalog_tb}
        \end{table*}

\subsection{Properties of the X-ray+OIR sources}
\label{xoir_properties}

\begin{figure}[!t]
        \centering
        \includegraphics[width=9cm]{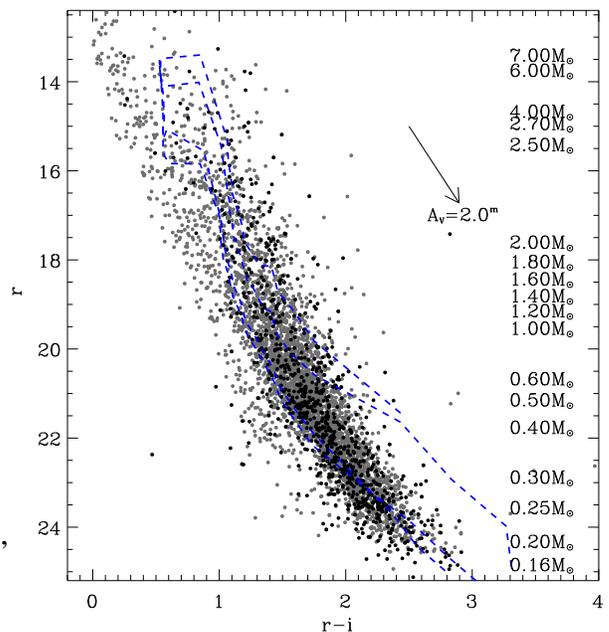}
        \caption{$r$ vs. $r-i$ diagrams with all the X-ray sources with a ``reliable'' OIR counterpart with good optical photometry marked with gray dots, and the Cyg~OB2 members with circumstellar disk marked with black dots. The isochrones, reddening vector, and masses are plotted as in Fig. \ref{colmag}}
        \label{colmag2}
        \end{figure}

\citet{GuarcelloDWD2013} studied the disk population of Cyg~OB2 within the field of the {\it Chandra} Cygnus~OB2 Legacy Survey. They selected and classified 1843 stars with disks associated with Cyg~OB2. Among these stars, a total of 444 have an X-ray counterpart: 368 class~II sources, 10 candidates with transition disks and 19 with pre-transition disks; 20 candidate accretors with intense H$\alpha$ emission; 6 blue stars with disks \citep{GuarcelloDMP2010}; and 16 candidate class~I sources. We also detected the X-ray emission from 52 known O stars, 57 known B stars, and 6 emission line objects selected by \citet{VinkDSW2008}. Other existing classifications of low-mass members of Cyg~OB2 made use of X-ray observations which are part of the Chandra Cygnus~OB2 Legacy Project \citep{AlbaceteColomboFMS2007,WrightDrake2009}, so that they are included in the list of candidate members produced by our survey. Fig.~\ref{colmag2} shows the $r$~vs.~$r-i$ diagram of the optical counterpart of the X-ray sources with ``reliable'' OIR matches. Their locus in this diagram is well delimited by the chosen isochrones, which correspond to the locus of the candidate members with circumstellar disks selected by \citet{GuarcelloDWD2013}. This suggests that this sample is dominated by young stars associated with Cyg~OB2. Optical-X-ray counterpart and disk-bearing sources show a slightly different distribution in $r$, being the former more numerous at bright magnitudes. This can be understood as a consequence of: 1) The fact that low-mass stars holds their inner disks for longer time than high-mass stars; 2) the X-ray catalog is not complete below 1$\,$M$_{\odot}$ \citep{Wrightinprep_sensitivity}; 3) the bright end of the $r$ distribution is more populated by candidate foreground stars detected in X-rays. For instance, a significant population of field stars apparently older than Cyg~OB2 stars lies in the bright blue part of the diagram (i.e.\ $r \leq 15.5^m$ and $r-i \leq 0.4^m$). These stars have an extinction significantly smaller than that of the Cyg~OB2 stars, as inferred from other color-color diagrams.  \par 

\begin{figure*}[!t]
        \centering
       \includegraphics[width=16cm]{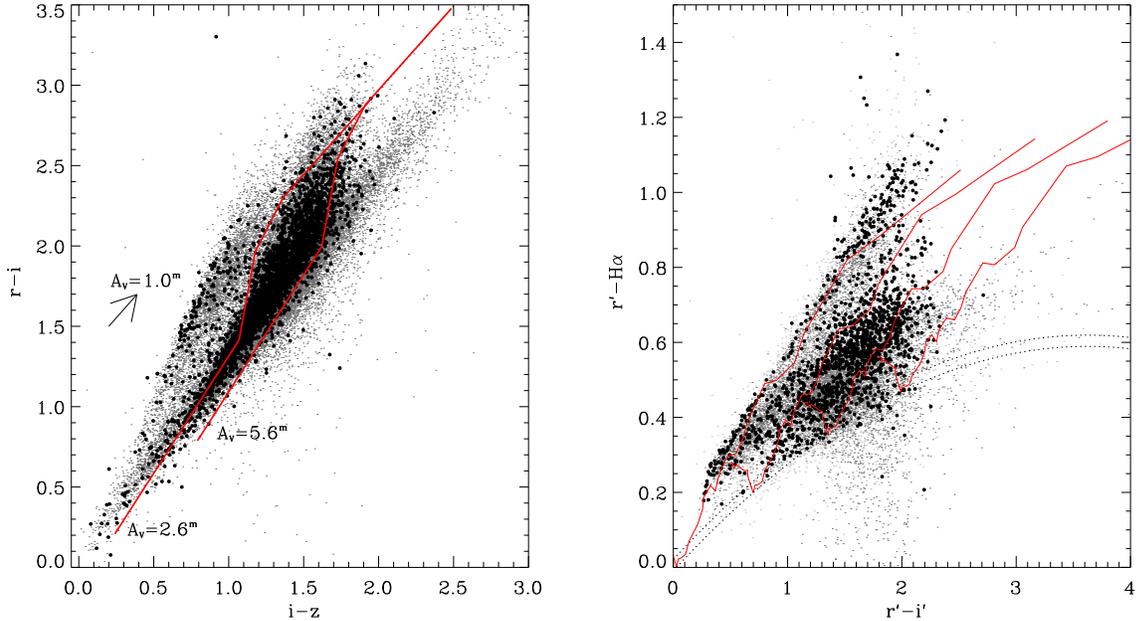}
        \caption{Optical color-color diagrams with all the sources with good photometry in the involved colors (gray dots) and the ``reliable'' counterparts of X-ray sources (black dots). In the $r-i$ vs. $i-z$ diagram the solid lines are $3.5\,$Myrs isochrones with $A_V=2.6^m$ and $A_V=5.6^m$; in the $r^{\prime}-H\alpha$ vs. $r^{\prime}-i^{\prime}$ the solid lines are ZAMS with increasing $E_{B-V}$ from 0 to 4.}
        \label{optplot}
        \end{figure*}

\begin{figure*}[!t]
        \centering
       \includegraphics[width=16cm]{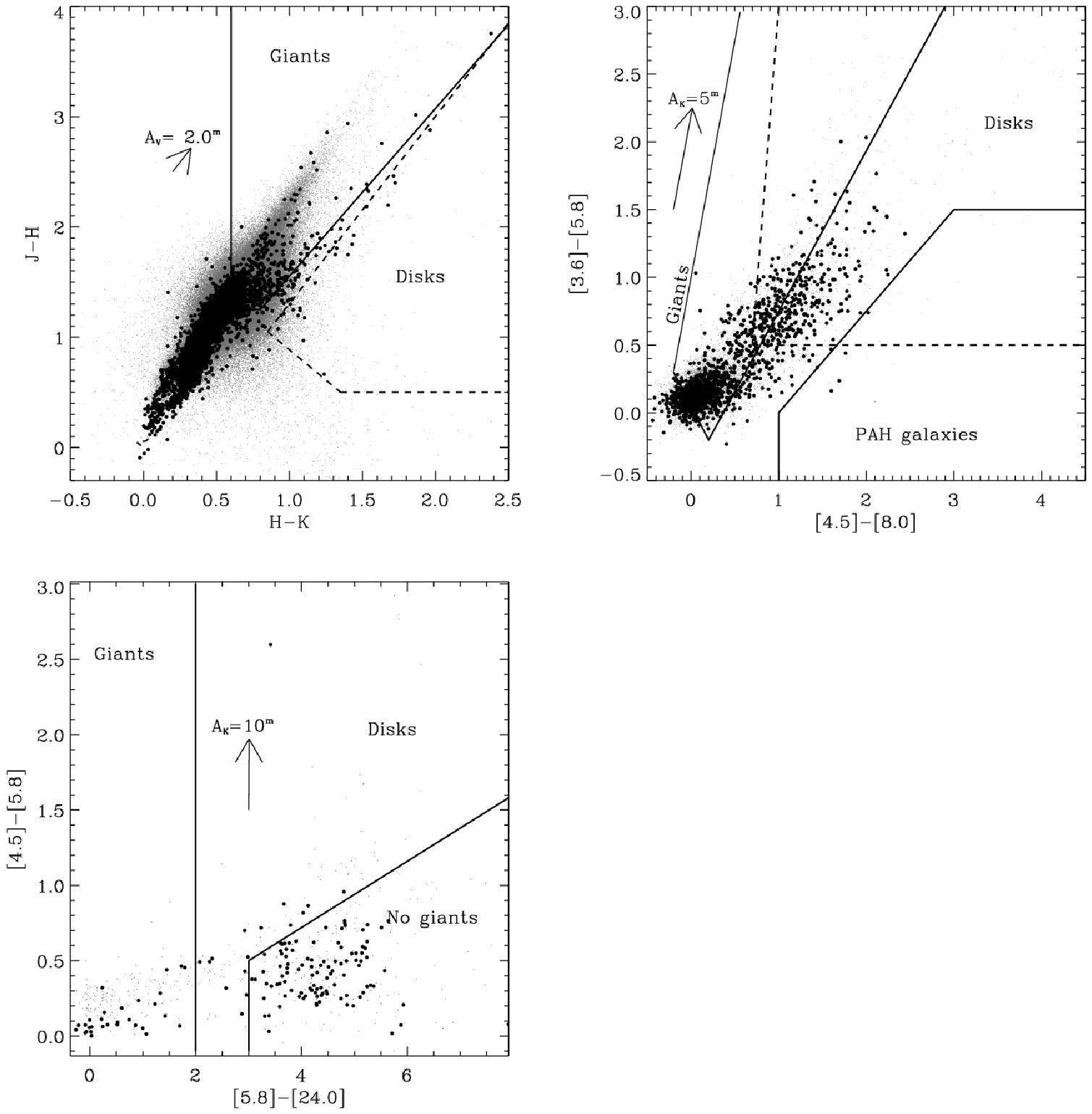}
        \caption{NIR color-color diagrams with all the sources with good photometry in the involved colors (gray dots) and the ``reliable'' counterparts of X-ray sources (black dots). The loci shown in the diagrams distinguish the typical colors expected from giants and normal stars at various extinction, disk-bearing stars, and background galaxies. In the right panel, giants can be found across all the diagram, except the locus marked with ``No giants''.}
        \label{nirplot}
        \end{figure*}

Figg.~\ref{optplot} and \ref{nirplot} show several optical and infrared color-color diagrams of the OIR sources with X-ray counterparts (black dots) classified as ``reliable matches'', together with all the sources meeting the criteria for good photometry in the relevant bands (i.e.\ errors in colors smaller than $0.15^m$). In the $r-i$ vs.\ $i-z$ diagram most of the X-ray sources with optical counterparts lie in the area delimited by the $3.5\,Myr$ isochrones from \citet{SiessDF2000} with extinction $A_V=2.6^m$ and $A_V=5.6^m$, which are the 10\%\ and 90\%\ quantiles, respectively,  of the optical extinction found in \citet{GuarcelloWDG2012}. The optical+X-ray sources to the left of the less extinguished isochrone are likely foreground sources, while possible background sources cannot be distinguished from the faintest stars associated with Cygnus~OB2. These properties are only weakly dependent on the isochrones used. \par

  Another diagram where the foreground population can be easily distinguished from Cyg~OB2 members is the $r^{\prime}-H\alpha$ vs.\ $r^{\prime}-i^{\prime}$ diagram. The solid lines are the ZAMS from \citet{DrewGIA2005} with $E_{B-V}=0^m, \, 1^m, \,2^m, \, 3^m$. The X-ray sources which lie close to the $E_{B-V}=0^m$ ZAMS are mainly in the foreground. Other classes of sources that can be distinguished are the background giants that mainly lie in the lower part of the diagram, below the ZAMS \citep{WrightGBD2008}, and accreting stars which show very red r-H$\alpha$ colors. Only a handful of IPHAS+X-ray sources lie in this part of the diagram. Moreover, candidate A stars are expected to populate the locus in this diagram within the dashed curved lines \citep{DrewGIS2008}.\par
  
The loci shown in the three infrared diagrams in Fig. \ref{nirplot} (i.e. the $Giants$, $Disk$, and $Galaxies$ loci) have been defined in \citet{GuarcelloDWD2013}. In the $J-H$ vs.\ $H-K$ diagram only a few NIR+X-ray sources lie in the $disk$ locus or at very high extinction. The first result is not surprising, since only 7.5\% of the selected stars with disks in Cyg~OB2 populate this locus \citep{GuarcelloDWD2013}; the second result suggests that the background contamination of the X-ray sources with NIR counterparts is low. Very small contamination from galaxies is also suggested by the $[3.6]-[5.8]$ vs. $[4.5]-[8.0]$ diagram, which is one of the diagrams used in \citet{GuarcelloDWD2013} for selecting disks and galaxies.  In this diagram, the population of X-ray sources with IR excesses due to the presence of a circumstellar disk lie inside and nearby the disk locus in the upper right part. Most of the X-ray sources with MIPS counterparts have intrinsic red colors, likely due to circumstellar disks, as shown in the $[4.5]-[5.8]$ vs. $[5.8]-[24]$ diagram. In all the discussed diagrams, the reddening vectors are taken from the extinction laws found by \citet{RiekeLebofsky1985,Donnell1994,FlahertyPMW2007}. 

\begin{figure}[!t]
        \centering
       \includegraphics[width=9cm]{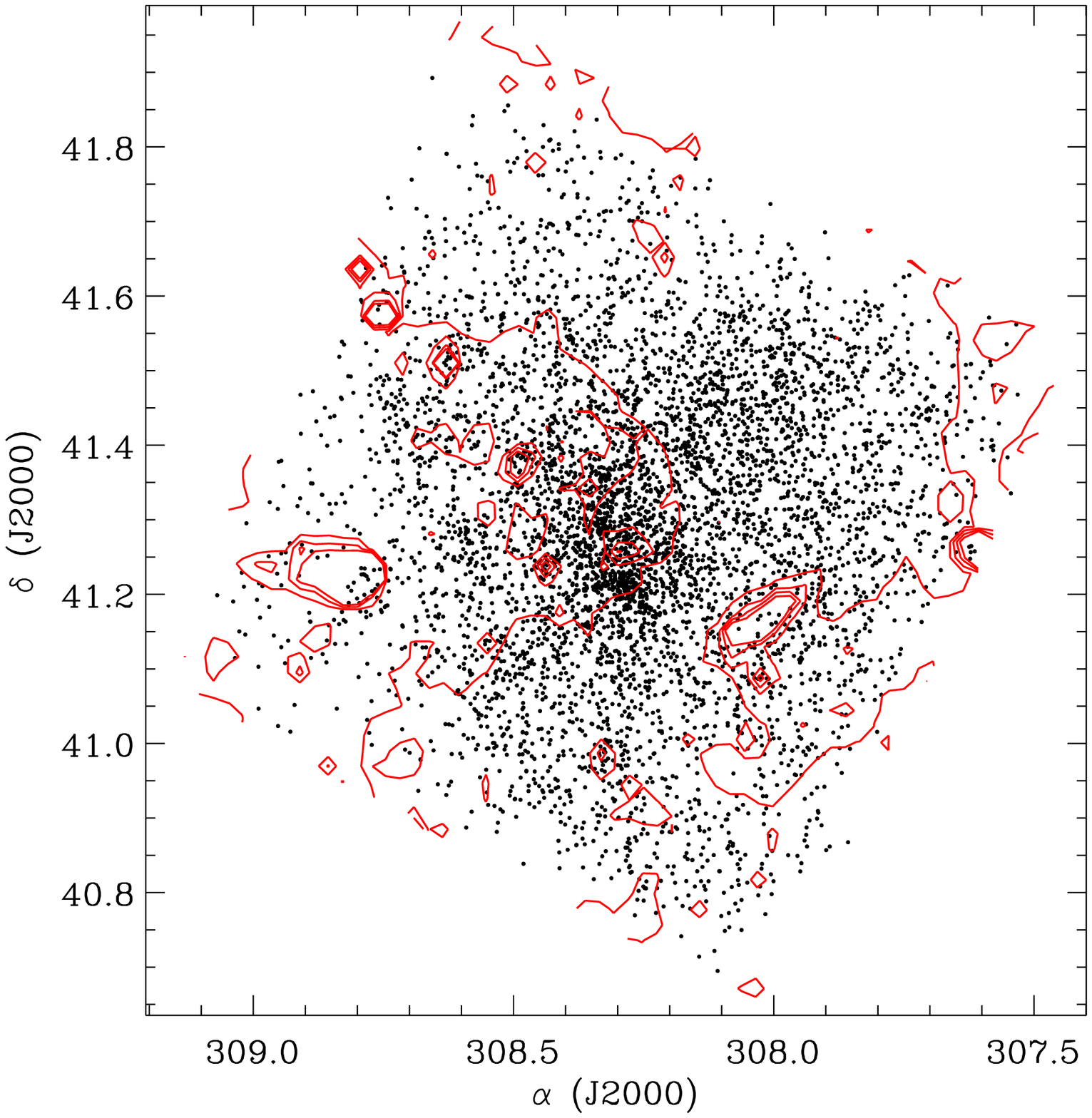}
        \caption{Spatial distribution of the X-ray sources with reliable OIR counterparts. The red solid lines mark the emission contours at $8.0\,\mu$m.}
        \label{spad_plot}
        \end{figure}

In Fig. \ref{spad_plot} we show the spatial distribution of all the X-ray sources with OIR counterparts, which are clearly clustered in the center of the field that roughly corresponds with the central cluster of Cyg~OB2 \citep{BicaBD2003,GuarcelloDWD2013}, but there is also a rich sparse population across the entire field. The contours mark the emission at $8.0\,\mu m$, showing the locations of the most dense nebular structures. \par

The analysis of the X-ray properties of the X-ray sources with OIR counterpart is beyond the scope of this paper, and it will be part of dedicated studies \citep{Flaccomioinprep1,Kashyapinprep}. A brief comparison of the observed X-ray photons energy between the X-ray sources with and without OIR counterpart is shown in the next section.

\subsection{Properties of the unmatched X-ray sources}
\label{onlyx_properties}

\begin{figure}[!h]
        \centering
        \includegraphics[width=9cm]{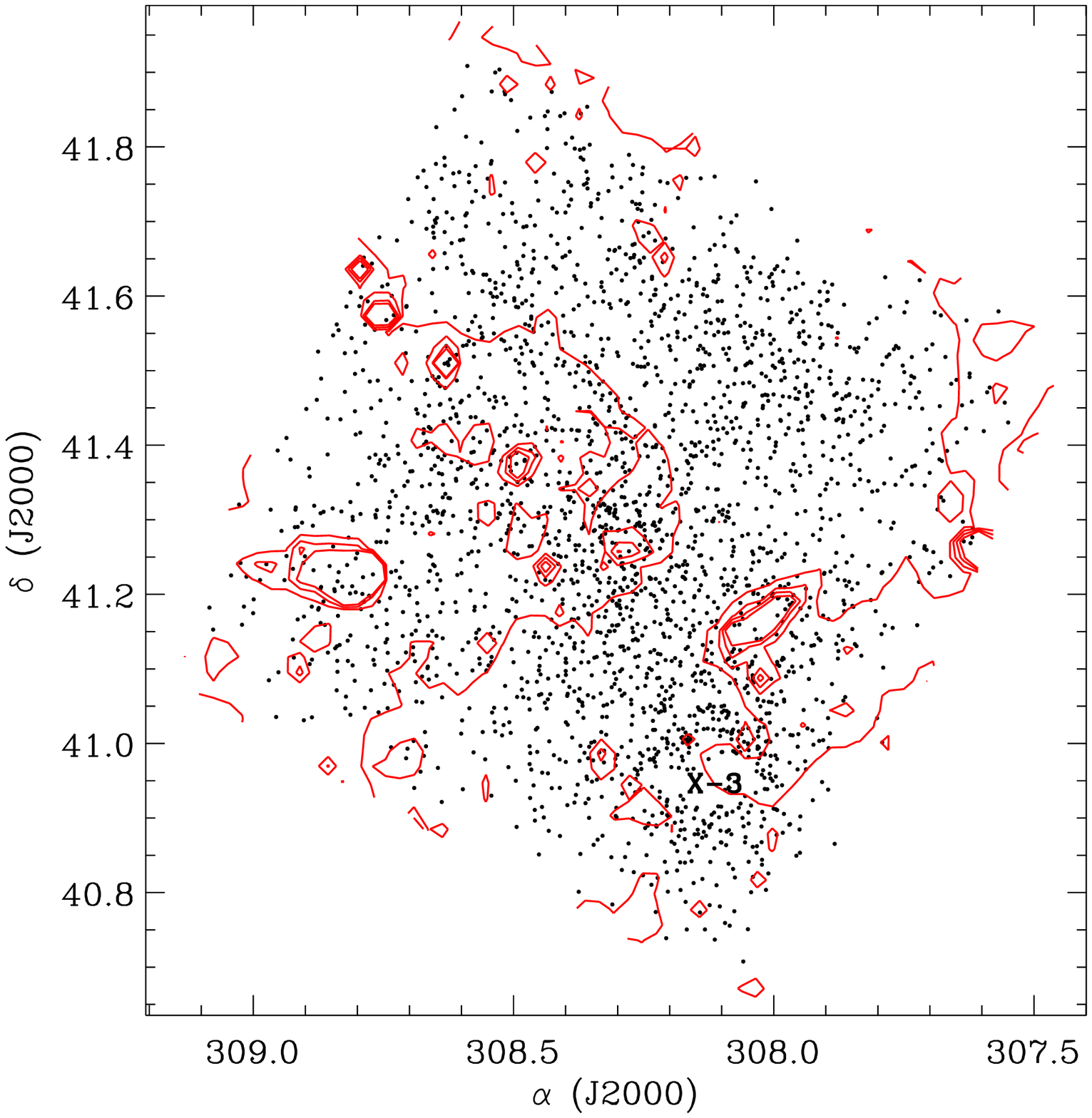}
        \caption{Spatial distribution of the X-ray sources without OIR counterparts, with overplotted the emission contours at $8.0\,\mu$m.}
        \label{spadis}
        \end{figure}

\begin{figure*}[!t]
        \centering
        \includegraphics[width=9cm]{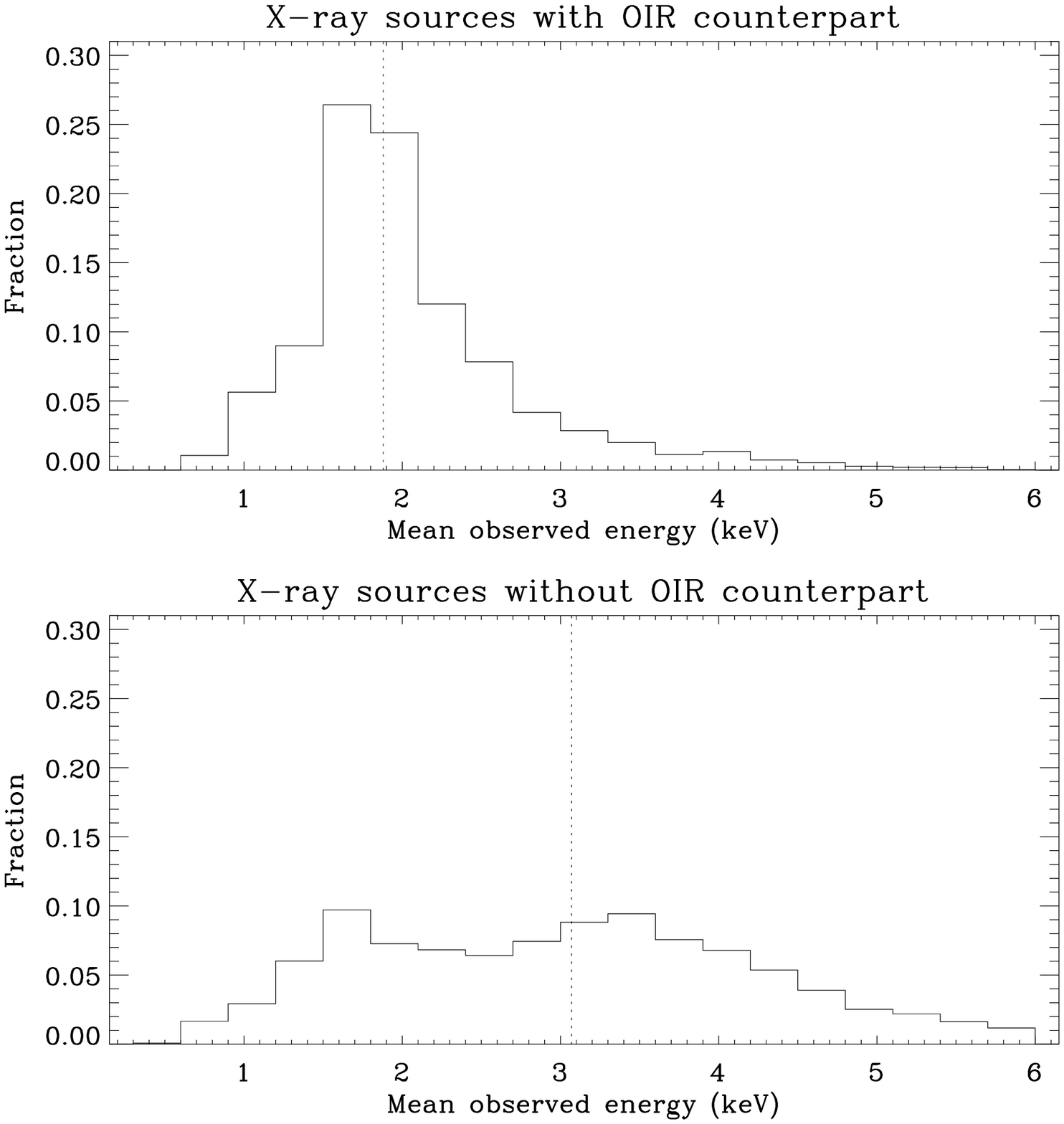}
        \includegraphics[width=9cm]{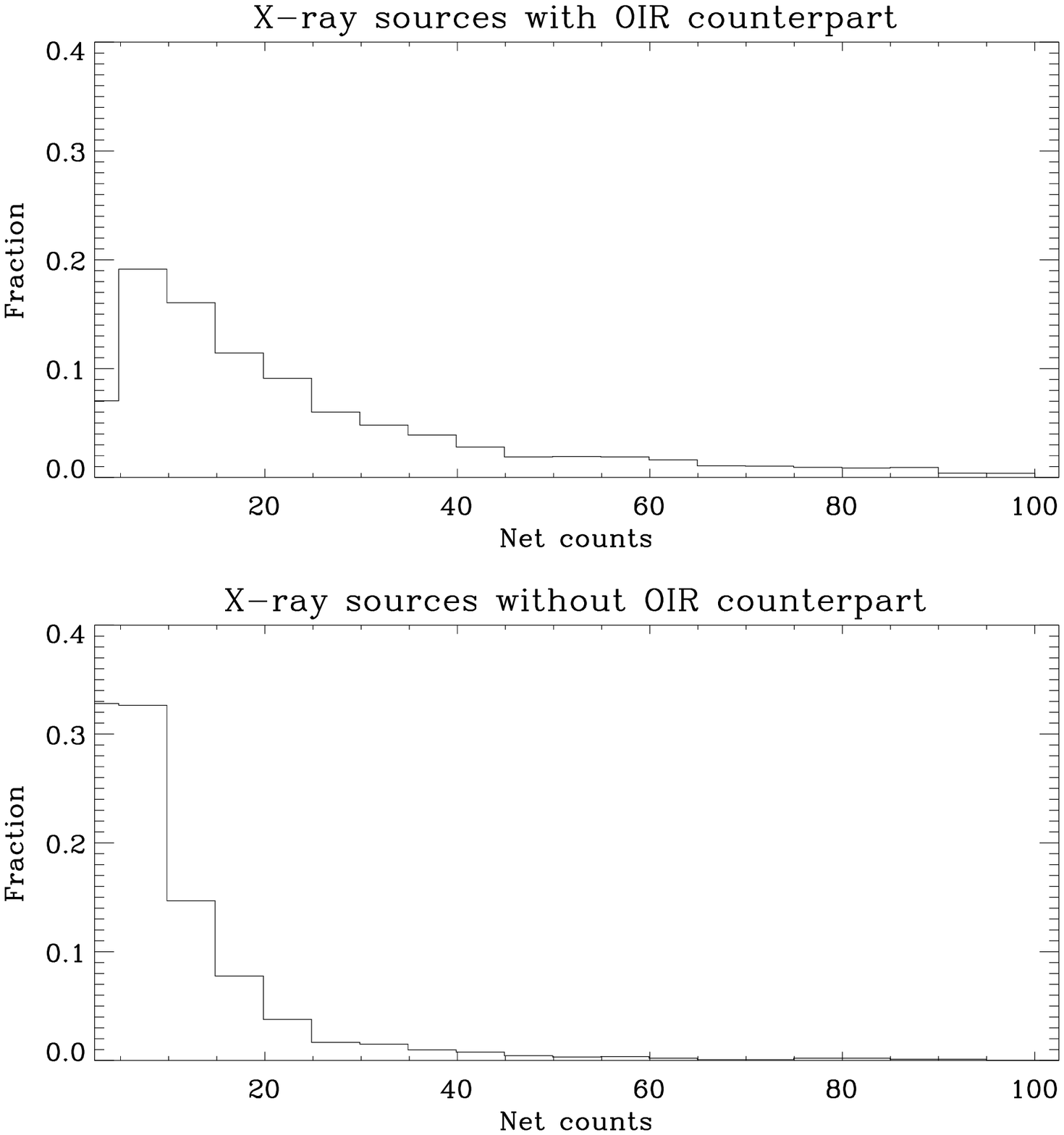}
        \caption{Distributions of the median photon energy (left panels) and net counts (right panels) of the X-ray sources with (upper panels) and without (lower panels) OIR counterparts. The vertical dotted lines in the left panels mark the median values of the distributions.}
        \label{xpro}
        \end{figure*}

The different nature of the X-ray sources with and without OIR counterparts is evident by comparing their spatial distribution and their median photon energy, shown in Figg. \ref{spadis} and \ref{xpro}. The distribution of the X-ray sources with OIR counterpart is peaked at about $1.65\pm0.15\,$keV. The distribution of the X-ray sources with no OIR counterpart is completely different, being flatter, shifted toward higher energies and approximately bimodal, with a high energy cut-off at about $3.45\pm0.15\,$keV. The two distributions can be understood if the former is dominated by stars associated with Cyg~OB2, while the latter is dominated by background and extragalactic sources, mainly active galactic nuclei, with also a small presence of possible less extinguished stars. The right panels show the distributions of the net counts for the two samples of X-ray sources. An evident excess of faint X-ray sources is observed among the X-ray sources with no OIR counterpart with compared to those with counterpart. \par

  Comparing the spatial distribution of the unmatched X-ray sources (Fig. \ref{spadis}) with that of the OIR+X-ray stars (bottom right panel in Fig. \ref{spad_plot}), it is evident that the latter show a high degree of clustering in the center of the field, as expected, while the former are almost uniformly distributed. For instance, in the central area, within $8^{\prime}$ from the median position of all the X-ray sources, fall the 24\% of the X-ray+OIR sources, and only the 13\% of the X-ray sources with no OIR counterpart.  In Fig. \ref{spadis} we do not observe a strong concentration of sources toward the most dense nebular structures, suggesting that the number of very extinguished members of Cyg~OB2 among the unmatched X-ray sources is low. We expect that a significant number of spurious X-ray detections are in the area around Cygnus~X-3, approximately at {the position marked as ``X-3'' in Fig. \ref{spadis}}. \par
  
\begin{figure}[!t]
        \centering
        \includegraphics[width=8.5cm]{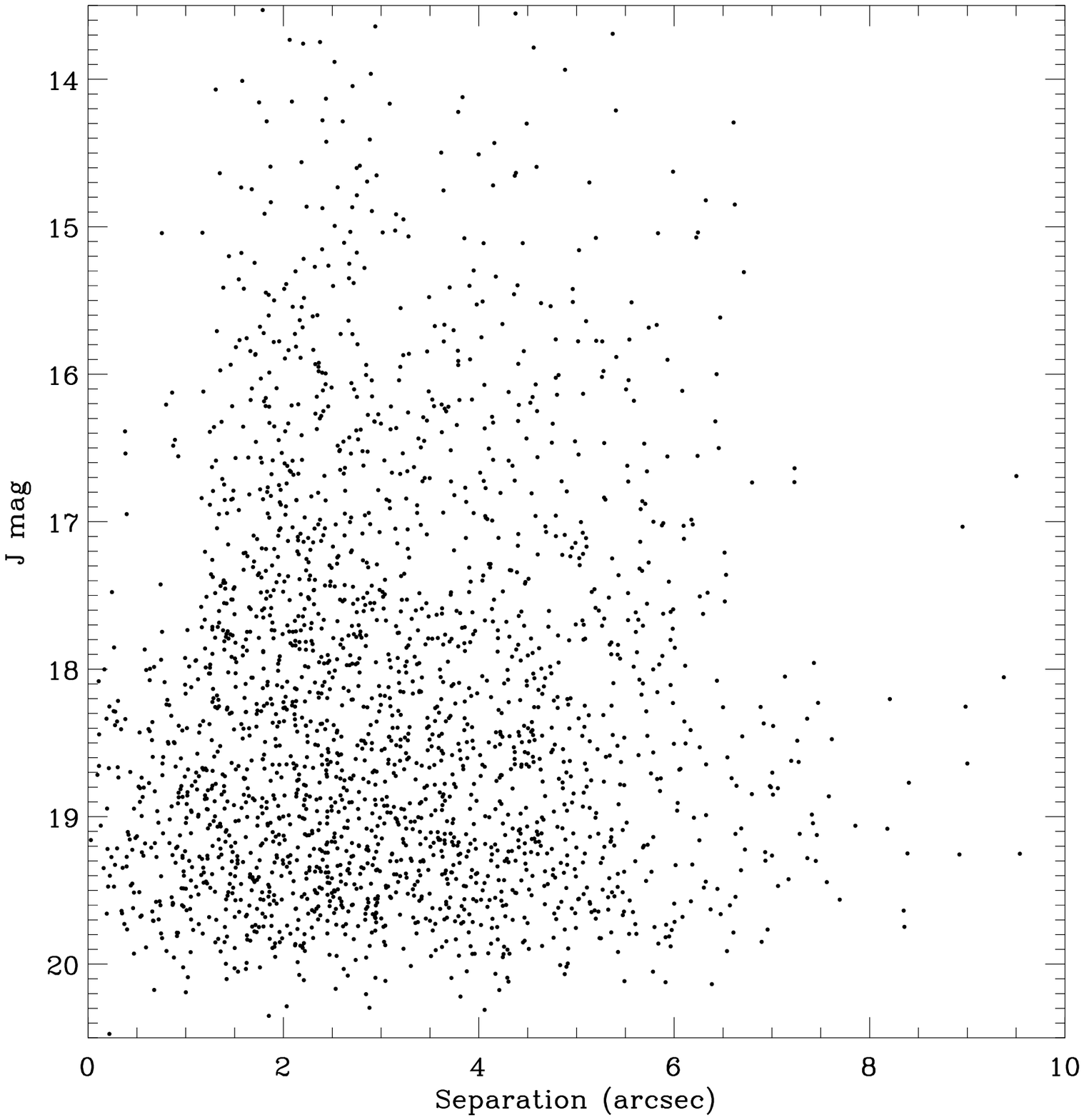}
        \caption{Distribution of the $J$ magnitude of the closest OIR sources to the unmatched X-ray sources vs. their angular separation.}
        \label{closOIR}
        \end{figure}

Fig.~\ref{closOIR} shows the diagram of the $J$ magnitudes of the closest OIR sources to each unmatched X-ray source versus their angular separation. As expected, the distribution of sources in this diagram is almost complementary to those in Fig.~\ref{checktay}. The vast majority of the sources here have large separations, with the exception of 206 X-ray sources with an OIR source closer than 1 arcsec.  Of these, 143 are faint in $J$ (having $J>18^m$), and only $18$ have a brighter $J$.  For 45 of these sources the $J$ photometry is absent or of  poor quality.  \par

\section{Conclusions}
\label{conclusions}

In this paper we compile a catalog of X-ray sources with optical and infrared counterparts of the massive star forming region Cygnus OB2. The X-ray catalog is obtained from the $1.08\,$Msec Chandra Cygnus OB2 Legacy Survey, covering an area of 1 square degree centered on Cyg~OB2.  The X-ray catalog contains 7924 sources. The optical-infrared catalogs have been compiled by merging several deep catalogs available for this region: From observations taken with OSIRIS@GTC (in $riz$ bands), the public catalogs SDSS/DR8 (in $ugriz$ bands), IPHAS/DR2 (in $r^{\prime}i^{\prime}H\alpha$ bands), UKIDSS/GPS and 2MASS/PSC ($JHK$ bands), and from the Spitzer Legacy Survey of the Cygnus~X region (in IRAC and $24.0\,\mu m$ bands). This catalog contains 329514 sources in the square degree area observed with Chandra/ACIS-I. 

  We discuss how a simple nearest-neighbor match would result in a highly contaminated catalog, with an excess of false positives and spurious matches. To avoid this, we adopted three different matching procedures, two based on a Maximum Likelihood approach plus the Bayesian method defined in \citet{NaylorBF2013}. The difference between the two Maximum Likelihood methods consist in how the expected magnitude distribution of the optical-infrared sources which are real counterparts of the X-ray sources. In one method this is obtained from a statistical approach from the observed magnitude distribution of the optical-infrared source nearby the X-ray positions. In the latter, with using an accurate closest-neighbor match.  All these three methods have been used and the results are merged in a single unique catalog. This merged catalog contains 5703 sources with X-ray and optical/infrared counterparts. We show that the most reliable optical-infrared counterparts are found with the Bayesian method and the Maximum Likelihood which uses the closest-neighbor match to define the expected correlated magnitude distribution, counting 5619 matches. The nature of these sources is discussed using optical and infrared color-color and color-magnitude diagrams: The vast majority are compatible with being associated with Cygnus~OB2; a low contamination from candidate background galaxies and giant stars is observed, while the foreground population looks to be more significantly represented.  Our combined catalog also contains 444 stars with disks, 52 O stars, and 57 B stars. 
 
\acknowledgments

This paper is based on an extensive dataset: optical data provided by observations made with the Gran Telescopio Canarias (GTC), installed in the Spanish Observatorio del Roque de los Muchachos of the Instituto de Astrof\'{i}sica de Canarias, in the island of La Palma; data from the IPHAS survey, based on observations carried out at the Isaac Newton Telescope, INT, operated on the island of La Palma by the Isaac Newton Group in the Spanish Observatorio del Roque de los Muchachos of the Instituto de Astrof\'{i}sica de Canarias; data from the SDSS Data Release 9, funded by the Alfred P. Sloan Foundation, the Participating Institutions, the National Science  Foundation, and the U.S. Department of Energy Office of Science; data products from the Two Micron All Sky Survey, which is a joint project of the University of Massachusetts and the Infrared Processing and Analysis Center/California Institute of Technology, funded by the National Aeronautics and Space Administration and the National Science Foundation; data based on observations made with the Spitzer Space Telescope, which is operated by the Jet Propulsion Laboratory, California Institute of Technology, under contract with NASA; and data obtained as part of the UKIRT Infrared Deep Sky Survey (UKIDSS), which used the UKIRT Wide Field Camera (WFCAM). The authors also made an extensive use of the software TOPCAT to manipulate the used catalogs. \par
M. G. G. acknowledges the grant PRIN-INAF 2012 (P.I. E. Flaccomio). N.J. W. acknowledges a Royal Astronomical Society Research Fellowship. They also have been supported by the Chandra grant GO0-11040X during the course of this work. J.J.D. was funded by NASA contract NAS8-03060 to the Chandra X-ray Center and thanks the Director, Belinda Wilkes, for continuing support. M. G. G. also acknowledges J. E. Drew for her important advices and help in the reduction and analysis of the OSIRIS and IPHAS data. 
\newpage

\addcontentsline{toc}{section}{\bf Bibliografia}
\bibliographystyle{aa}
\bibliography{biblio}

\end{document}